\documentclass[letterpaper, 12pt]{article}
\usepackage{jheppub}

\usepackage{graphicx}
\usepackage{bbm}
\usepackage{float}
\usepackage{amssymb}
\usepackage{amsthm}
\usepackage{multirow}
\usepackage{physics}
\usepackage{afterpage}
\usepackage{rotating}
\usepackage{upgreek}
\usepackage{amsmath}
\usepackage{mathtools}
\usepackage{epstopdf}
\usepackage{mathrsfs}
\usepackage[labelformat=simple]{subcaption}
\usepackage{calligra}
\usepackage[T1]{fontenc}  

\begin{document}

\begin{titlepage}

\setcounter{page}{1} \baselineskip=15.5pt \thispagestyle{empty}
{\flushright {YNU-HEP-26-062}\\}
		
\bigskip
		
\vspace{1.4cm}
\begin{center}
{\Large \bfseries Multi-Component Axion Dark Matter Revisited\vspace{0.24cm}\\ and the Role of Ultralight ALPs}
\end{center}
\vspace{0.15cm}
			
\begin{center}
{\fontsize{14}{30}\selectfont Hai-Jun Li}
\end{center}
\begin{center}
\vspace{0.25 cm}
\textsl{Department of Physics, Yunnan University, Kunming 650091, China}\\

\vspace{-0.1 cm}				
\begin{center}
{E-mail: \textcolor{blue}{\tt {lihaijun@ynu.edu.cn}}}
\end{center}	
\end{center}
\vspace{0.6cm}
\noindent

In this work, we revisit the multi-component axion dark matter (DM) scenario and explore the role of ultralight axion-like particles (ALPs). 
After briefly reviewing maximal axion mixing, we incorporate a stochastic axion mixing mechanism that generalizes it by relaxing the requirement of a uniform hierarchy among the axion decay constants. 
We find that, depending on the parameters, either the QCD axion or an ALP can dominate the DM abundance after mixing; once axion mass contributions are included, the heaviest ALPs tend to prevail. 
In a two-component framework, stochastic mixing significantly enlarges the allowed parameter space for the QCD axion decay constant, enabling the QCD axion to constitute the entirety of cold DM. 
The corresponding ALP mass range remains largely unconstrained by current experiments and is within reach of future searches. 
In the multi-component case, a larger ALP population leads to the emergence of more ultralight ALPs with large decay constants.
These ultralight ALPs represent compelling targets for upcoming axion searches.
 		
\vspace{3cm}
			
\bigskip
\noindent\today
\end{titlepage}
			
\setcounter{tocdepth}{2}
			
\hrule
\tableofcontents			
\bigskip\medskip
\hrule
\bigskip\bigskip
 

\section{Introduction}

Despite substantial observational evidence for dark matter (DM), its true nature remains unknown, making the axion a leading DM candidate \cite{Cirelli:2024ssz}.
The axion, commonly identified as the QCD axion, was originally postulated through the Peccei-Quinn (PQ) mechanism --- known as the Peccei-Quinn-Weinberg-Wilczek (PQWW) model --- to dynamically address the strong CP problem in the Standard Model \cite{Peccei:1977hh, Peccei:1977ur, Weinberg:1977ma, Wilczek:1977pj}.
The two most prominent models for the QCD axion are the Kim-Shifman-Vainshtein-Zakharov (KSVZ) and Dine-Fischler-Srednicki-Zhitnitsky (DFSZ) models \cite{Kim:1979if, Shifman:1979if, Dine:1981rt, Zhitnitsky:1980tq}.

Beyond the QCD axion, a vast array of ultralight axion-like particles (ALPs) can arise from higher-dimensional gauge fields, a category that naturally encompasses the QCD axion candidate \cite{Witten:1984dg, Green:1984sg, Choi:2003wr, Reece:2024wrn}. It is well established that string theory anticipates a multitude of such axions \cite{Svrcek:2006yi, Conlon:2006tq}, a concept collectively known as the string axiverse \cite{Arvanitaki:2009fg, Acharya:2010zx, Cicoli:2012sz}.
Currently, both QCD axions and ALPs have attracted significant interest as prime candidates for cold DM, owing to their non-thermal production in the early Universe through the misalignment mechanism \cite{Preskill:1982cy, Abbott:1982af, Dine:1982ah}. 
The oscillation of a coherent axion field within its potential well contributes substantially to the total DM density. 
For comprehensive reviews, see $\rm e.g.$ refs.~\cite{Marsh:2015xka, OHare:2024nmr}.

Regarding QCD axion DM, if one assumes an initial misalignment angle of order unity, the standard misalignment mechanism sets an upper bound on the classical QCD axion window at $10^{11}-10^{12}\, {\rm GeV}$. 
Consequently, a major challenge for QCD axion DM is the overproduction problem, which occurs if the decay constant exceeds $10^{12}\, {\rm GeV}$. 
Conversely, in the recently proposed dark dimension scenario \cite{Montero:2022prj}, where the decay constant is restricted to a narrow band of $10^{9}-10^{10}\, \rm GeV$, the QCD axion faces the opposite issue of insufficient abundance \cite{Gendler:2024gdo, Li:2024jko}. 
Fortunately, axion mixing provides a natural resolution to both dilemmas. 
In the light QCD axion scenario \cite{Daido:2015cba, Li:2023uvt}, the relic density can be efficiently suppressed, whereas in the heavy QCD axion scenario \cite{Cyncynates:2023esj}, it can be considerably enhanced. 
In this context, axion mixing typically denotes the mass mixing between the QCD axion and an ALP \cite{Hill:1988bu, Ho:2018qur, Li:2023xkn, Li:2025imm, Murai:2024nsp, Li:2025uwq, Murai:2025wbg}.

Along these lines, we introduced a multi-component axion DM scenario in ref.~\cite{Li:2025uwq}, where the cold DM consists of a combination of the QCD axion and many ultralight ALPs through maximal axion mixing \cite{Li:2025cep}. 
We find that in the light axion scenario, the energy density is typically dominated by the lightest ALP after mixing.
Conversely, in the heavy axion scenario, the energy density may be governed by either the QCD axion or the non-lightest ALPs, depending on the ALP decay constants. 
Under specific circumstances, the QCD axion can account for the majority of the DM budget.
Furthermore, in ref.~\cite{Li:2026rga}, we proposed a novel stochastic axion mixing mechanism, which represents a more general framework as it does not rely on restrictive conditions for the axion decay constants.
See also ref.~\cite{Asadi:2025cvm, deGiorgi:2025ldc, Baryakhtar:2026oun, FernandezNavarro:2026cyu, Gorgulho:2026psp, Langhoff:2026qqg, Sengupta:2026gyg, Lei:2026cbn} for recent multi-axion frameworks.

In this work, we revisit the multi-component axion DM scenario and explore the role of ultralight ALPs within this framework.
We begin by briefly reviewing this scenario, which corresponds to maximal axion mixing.
Subsequently, we introduce the recently proposed stochastic mixing mechanism and derive several new conclusions.
In the absence of axion mass contributions, we find that either the QCD axion or any of the ALPs can become dominant after mixing.
Once axion mass contributions are included, the ratio $\rho_{A_i,0}/\rho_{a_0,0}$ increases with the ALP mass, implying that the heaviest ALPs tend to prevail.
Next, we investigate the role of ultralight ALPs within the multi-component scenario under the stochastic mixing mechanism.
In a two-component framework, we find that the parameter space for the QCD axion decay constant --- allowing the QCD axion to constitute the entirety of cold DM --- is significantly enlarged.
The corresponding ALP mass range remains largely unconstrained and is accessible to future detection experiments.
Extending to the multi-component case, we find that a larger ALP population in the stochastic mixing model leads to the emergence of more ultralight ALPs, thus allowing for a greater number of viable ALPs with large decay constants.
These ultralight ALPs, which currently evade experimental constraints, constitute compelling targets for future axion searches.

The rest of this paper is structured as follows. 
Section~\ref{multi_component} briefly reviews the multi-component axion DM scenario with maximal axion mixing.
In section~\ref{stochastic_axion_mixing}, we extend this scenario by incorporating the stochastic mixing mechanism. 
Section~\ref{ultra_light_ALPs} then investigates the role of ultralight ALPs within this framework, taking stochastic mixing into account. 
Finally, we conclude the paper in section~\ref{sec_Conclusion}. 
  
\section{Overview of multi-component axion dark matter}
\label{multi_component}   
  
In this section, we briefly review the recently proposed multi-component axion DM scenario \cite{Li:2025uwq}.
In this framework, the DM is composed of the QCD axion ($a_0$) and $N$ light ALPs $A_i$ ($i=1,\cdots,N$).
Note that here we can assume the axion mass ordering to be $m_{A_1}<\cdots<m_{A_N}<m_{a_0,0}$.
The low-energy effective potential for the multi-component axion DM model is given by
\begin{eqnarray}
V_{\rm eff}=\sum_{i=0}^N\Lambda_i^4\left[1-\cos\left(\sum_{j=0}^N n_{ij} \theta_j \right)\right]\, ,
\label{potential}
\end{eqnarray} 
where $\Lambda_i$, $\theta_i$, and $n_{ij}$ represent the overall scales, the axion angles, and the domain wall numbers, respectively.
In this scenario, the present total axion energy density (with the current CMB temperature $T_0$) is given by
\begin{eqnarray}
\rho_{\rm tot,0}=\rho_{a_0,0}+\sum_{i=1}^N \rho_{A_i,0}\, ,
\end{eqnarray} 
where $\rho_{a_0,0}$ and $\rho_{A_i,0}$ are the axion energy densities after taking into account the axion mixing effect.
Specifically, the present energy density of the QCD axion and ALPs after mixing can be respectively expressed as
\begin{eqnarray}
\text{QCD axion:}~\rho_{a_0,0}&=&\dfrac{1}{2} m_{a_0,0} m_{A_N}  f_{A_N}^2 \Theta_{{\rm i},A_N}^2 \left(\dfrac{\mathfrak{a}_{{\rm osc},A_N}}{\mathfrak{a}_0}\right)^3\, ,\\
\text{ALPs:}~\rho_{A_i,0}&=&
\begin{cases}
\dfrac{1}{2} m_{A_i} m_{a_0,{\rm osc}} f_{a_0}^2 \theta_{{\rm i},a_0}^2 \left(\dfrac{\mathfrak{a}_{{\rm osc},a_0}}{\mathfrak{a}_0}\right)^3\, , &\text{if } i=1\\
\dfrac{1}{2}m_{A_{i-1}} m_{A_i} f_{A_{i-1}}^2 \Theta_{{\rm i},A_{i-1}}^2 \left(\dfrac{\mathfrak{a}_{{\rm osc},A_{i-1}}}{\mathfrak{a}_0}\right)^3 \, , &\text{if } 1<i\leq N
\end{cases}
\end{eqnarray} 
where $m_{a_0}$ and $m_{A_i}$ are the masses of the QCD axion and ALPs, respectively, $f_{a_0}$ and $f_{A_i}$ are the decay constants, $\theta_{{\rm i},a_0}$ and $\Theta_{{\rm i},A_i}$ are the initial misalignment angles, $\mathfrak{a}$ is the scale factor, and ``osc$"$ corresponds to the oscillation temperature $T_{\rm osc}$.
Then we obtain the ratios of the energy densities between the QCD axion and ALPs \cite{Li:2025uwq} 
\begin{eqnarray} 
\dfrac{\rho_{a_0,0}}{\rho_{A_i,0}}\simeq
\begin{cases}
\dfrac{m_{a_0,0} \sqrt{m_{a_0,{\rm osc}}}}{m_{A_i} \sqrt{m_{A_N}}} \left(\dfrac{f_{A_N}}{f_{a_0}}\right)^2\, , &\text{if } i=1\\
\dfrac{m_{a_0,0} \sqrt{m_{A_{i-1}}}}{m_{A_i}\sqrt{m_{A_N}}} \left(\dfrac{f_{A_N}}{f_{A_{i-1}}}\right)^2\, , &\text{if } 1<i\leq N
\end{cases}
\label{ratio}
\end{eqnarray}  
where the initial misalignment angles are assumed to be of order one.  

The matrix of domain wall numbers $n_{ij}$ in eq.~\eqref{potential} is taken as a $(N+1)\times(N+1)$ matrix $\mathfrak{n}_{\mathfrak{ij}}$, beginning with indices $\mathfrak{i}=0$ and $\mathfrak{j}=0$.
In ref.~\cite{Li:2025uwq}, the maximal mixing is adopted for the axion mixing mechanism.
Therefore, the matrices of domain wall numbers $n_{ij}$ are described as
\begin{eqnarray}
\mathfrak{n}_{\mathfrak{ij}}=
\left(
\begin{array}{cccccc}
1  & ~ 0 & ~ 0 & ~ 0 &~\cdots & ~ 0\\
1  & ~ 1 & ~ 0 & ~ 0 &~\cdots & ~ 0\\
1  & ~ 0 & ~ 1 & ~ 0 &~\cdots & ~ 0\\
1  & ~ 0 & ~ 0 & ~ 1 &~\cdots & ~ 0\\
\vdots  & ~\vdots & ~\vdots &~\vdots & ~\ddots & ~ \vdots\\
1  & ~ 0 & ~ 0 & ~ 0 & ~ \cdots & ~ 1
\end{array}
\right)\, , \quad
\mathfrak{n}_{\mathfrak{ij}}=
\left(
\begin{array}{cccccc}
1  & ~ 1 & ~ 1 & ~ 1 &~\cdots & ~ 1\\
0  & ~ 1 & ~ 0 & ~ 0 &~\cdots & ~ 0\\
0  & ~ 0 & ~ 1 & ~ 0 &~\cdots & ~ 0\\
0  & ~ 0 & ~ 0 & ~ 1 &~\cdots & ~ 0\\
\vdots  & ~\vdots & ~\vdots &~\vdots & ~\ddots & ~ \vdots\\
0  & ~ 0 & ~ 0 & ~ 0 & ~ \cdots & ~ 1
\end{array}
\right)\, .
\label{nij_max}
\end{eqnarray}
The former corresponds to the light QCD axion scenario, where the decay constants of all ALPs are smaller than that of the QCD axion, $f_{A_i}\ll f_{a_0}$; the latter corresponds to the heavy QCD axion scenario, where the decay constants of all ALPs are larger than that of the QCD axion, $f_{A_i}\gg f_{a_0}$.
Then, through eq.~\eqref{ratio}, we can observe that different axions will become dominant after mixing:
\begin{eqnarray}
&&\text{Light axion scenario:}~\text{the lightest ALP $A_1$ dominates},\\ 
&&\text{Heavy axion scenario:}~\text{the QCD axion and ALPs (except $A_1$) may dominate}.~~~
\end{eqnarray}
In the light axion scenario, only the ‌lightest‌ ALP, denoted as $A_1$, plays a dominant role.
In contrast, in the heavy axion scenario, both the QCD axion and the remaining ALPs (except $A_1$) have the potential to become dominant. 
This outcome depends on the relative values of the ALP decay constants.
If all ALP decay constants are very similar, the QCD axion and other non-lightest ALPs will be dominant.
However, if there are significant differences among the ALP decay constants, different axions may dominate, necessitating a detailed analysis based on specific model parameters.

The maximal mixing mechanism is adopted here; readers are referred to ref.~\cite{Li:2025uwq} for details on the associated energy density transfer. Next, we extend our analysis of the multi-component axion DM scenario by incorporating the recently proposed stochastic mixing mechanism.

\section{New perspectives with stochastic axion mixing}
\label{stochastic_axion_mixing}   
 
In this section, we extend the multi-component axion DM scenario by incorporating the stochastic axion mixing mechanism \cite{Li:2026rga}.
This mechanism is essential for making contact with more general ultraviolet constructions and for exploring the resulting axion energy densities in a unified way. 
 
We note that for maximal axion mixing to occur, the decay constants of the ALPs must be either much smaller or much larger than that of the QCD axion. 
In the previous section, we assumed that all $N$ ALPs satisfy this condition. 
Otherwise, the effective number of mixings would be reduced, $\rm i.e.$, less than $N$. 
However, in the stochastic axion mixing scenario, even if some ALPs have decay constants smaller than that of the QCD axion while others have larger ones, the total number of mass crossings remains $N$. 
Therefore, stochastic mixing represents a more general mixing mechanism, as it does not depend on restrictive conditions for axion decay constants. 
It should be noted that, regardless of the mixing mechanism, the masses of these ALPs must all be smaller than the zero-temperature mass of the QCD axion and must be non-degenerate and well separated from one another.

In the stochastic axion mixing scenario, the matrix of domain wall numbers $n_{ij}$ in eq.~\eqref{potential} is given by
\begin{eqnarray}
\mathfrak{n}_{\mathfrak{ij}}=
\left(
\begin{array}{cccccc}
1  & ~ d_1 & ~ d_2 & ~ d_3 &~\cdots & ~ d_N\\
\tilde{d}_1 & ~ 1 & ~ 0 & ~ 0 &~\cdots & ~ 0\\
\tilde{d}_2  & ~ 0 & ~ 1 & ~ 0 &~\cdots & ~ 0\\
\tilde{d}_3  & ~ 0 & ~ 0 & ~ 1 &~\cdots & ~ 0\\
\vdots  & ~\vdots & ~\vdots &~\vdots & ~\ddots & ~ \vdots\\
\tilde{d}_N  & ~ 0 & ~ 0 & ~ 0 & ~ \cdots & ~ 1
\end{array}
\right)\, ,
\label{nij_new}
\end{eqnarray}
where $\tilde{d}_i=1-d_i$, and the $d_i$ are defined by
\begin{eqnarray}
d_i=
\begin{cases}
0\,, &\text{if } f_{A_i} \ll f_{a_0}\\
1\,, &\text{if } f_{A_i} \gg f_{a_0}
\end{cases}
\end{eqnarray}
corresponding to the light and heavy axion scenarios discussed above, respectively.
Thus, each $d_i$ selects whether the $i$-th ALP lies in the light or heavy regime
relative to the QCD axion, and the matrix in eq.~\eqref{nij_new} encodes all such
choices simultaneously.
We see that the matrix of domain wall numbers has a single unified form here, as it combines the two previous scenarios into one.
Although a duality between the light and heavy axion scenarios can be explicitly demonstrated in the two-axion mixing case, unifying the two scenarios within a single framework via stochastic mixing represents a novel approach to multi-axion mixing.

\begin{figure}[t]
\centering
\includegraphics[width=0.7\textwidth]{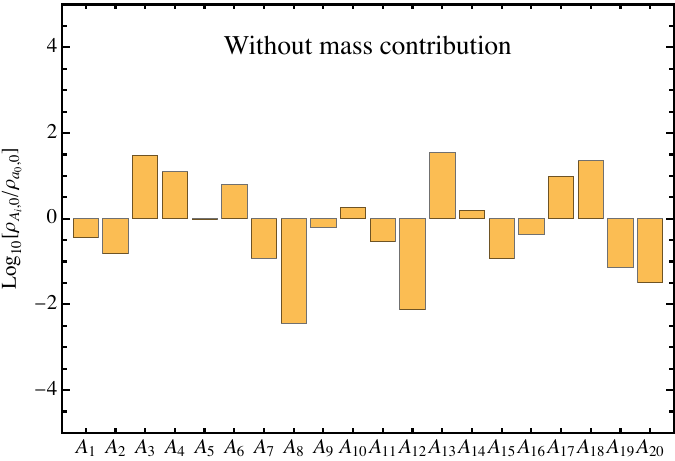}
\includegraphics[width=0.7\textwidth]{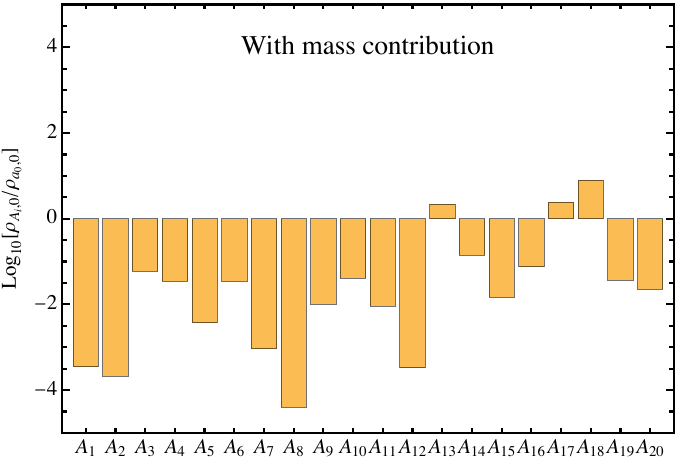}
\caption{The ratios of the energy densities $\rho_{A_i,0}/\rho_{a_0,0}$ in the stochastic mixing scenario.
‌The upper and lower panels represent the cases without and with axion mass contributions, respectively.
Here we set $f_{a_0}=10^{12}\, {\rm GeV}$, and the $f_{A_i}$ are chosen as 20 random values in the range $10^{11}-10^{13}\, {\rm GeV}$.}
\label{fig_ratio_stochastic_a}
\end{figure} 

\begin{figure}[t]
\centering
\includegraphics[width=0.7\textwidth]{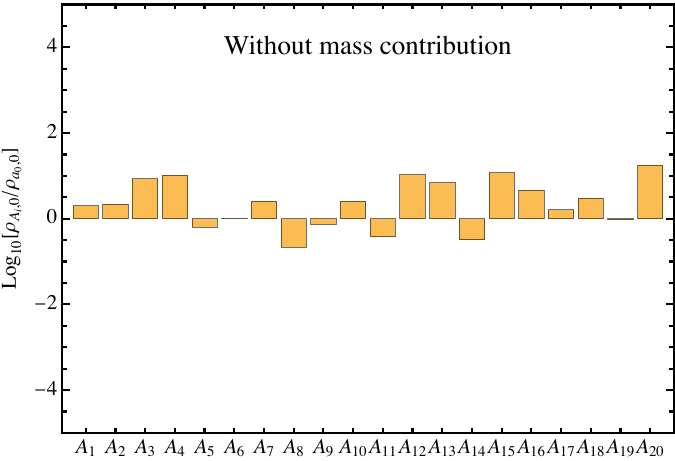}
\includegraphics[width=0.7\textwidth]{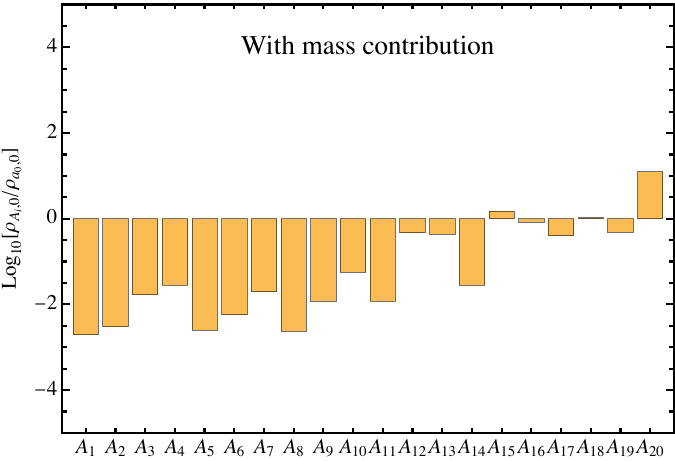}
\caption{Same as figure~\ref{fig_ratio_stochastic_a} but for narrower intervals of $f_{A_i}$.
Here we set $f_{a_0}=10^{12}\, {\rm GeV}$, and the $f_{A_i}$ are chosen as 20 random values in the range $10^{11.5}-10^{12.5}\, {\rm GeV}$.}
\label{fig_ratio_stochastic_b}
\end{figure} 

\begin{figure}[t]
\centering
\includegraphics[width=0.7\textwidth]{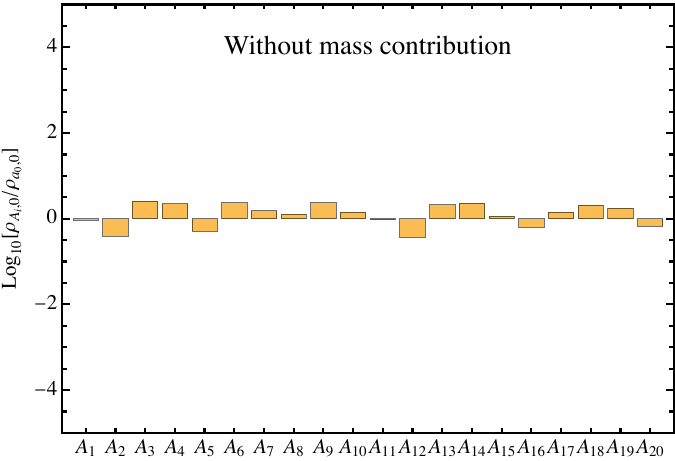}
\includegraphics[width=0.7\textwidth]{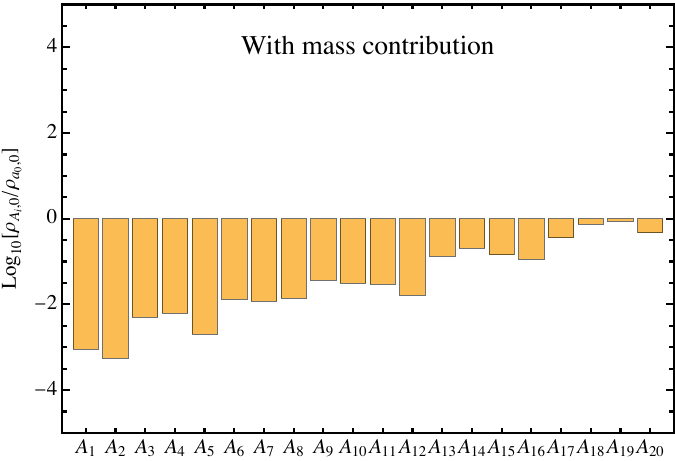}
\caption{Same as figure~\ref{fig_ratio_stochastic_a} but for narrower intervals of $f_{A_i}$.
Here we set $f_{a_0}=10^{12}\, {\rm GeV}$, and the $f_{A_i}$ are chosen as 20 random values in the range $10^{11.75}-10^{12.25}\, {\rm GeV}$.}
\label{fig_ratio_stochastic_c}
\end{figure} 

We now incorporate the stochastic mixing mechanism into the multi-component axion DM scenario. 
We find that the evolution of the axion field and the transfer of axion energy density resemble those discussed in the case of maximal mixing in ref.~\cite{Li:2025uwq}. 
However, there are two main differences.
The first lies in the number of axion mixings: as mentioned earlier, for a system of $N$ ALPs, the number of mixings remains $N$. 
The second is the intrinsic difference in axion decay constants between the two scenarios, which results in distinct axion energy densities.

In figure~\ref{fig_ratio_stochastic_a}, we display the distribution of axion energy density ratios in the stochastic mixing scenario, considering a system with one QCD axion and twenty ALPs. 
The upper panel illustrates the scenario neglecting axion mass contributions, thereby showing only the effects of axion decay constants. 
We observe that in this case, either the QCD axion or any of the ALPs can dominate after mixing.
This contrasts with our previous discussion of maximal mixing in ref.~\cite{Li:2025uwq}, primarily due to the distinct intrinsic choices of decay constants in the range of $f_{A_i}\sim 10^{11}-10^{13}\, {\rm GeV}$.
Furthermore, new insights emerge when we account for axion mass contributions. 
In the light and heavy axion scenarios, the spectral signature of axion masses has been discussed previously. 
In fact, this spectral signature, $\rm i.e.$, the hierarchical distribution of axion masses, applies equally to the stochastic axion mixing, which can be characterized by
\begin{eqnarray}
\dfrac{m_{A_i}}{m_{A_{i+1}}} \lesssim \dfrac{1}{2} \, , \quad \dfrac{m_{A_i}}{m_{a_0,0}} \lesssim \left(\dfrac{1}{2}\right)^{N-i+1}\, ,
\label{mass_condition}
\end{eqnarray}
where $N$ represents the number of ALPs. 
These inequalities represent the mass hierarchy required for efficient mass crossing and are understood for decay constants of the same order of magnitude; see ref.~\cite{Li:2025tdi} for further details.
In the lower panel of figure~\ref{fig_ratio_stochastic_a}, we present the distribution of axion energy density ratios incorporating axion mass contributions. 
We observe that the ratio increases with the ALP mass, implying that the heaviest ALPs dominate. 

The aforementioned conclusions become more evident when the ALP decay constants are drawn from a narrower range, as shown in figures~\ref{fig_ratio_stochastic_b} and \ref{fig_ratio_stochastic_c} for the ranges of $f_{A_i}\sim 10^{11.5}-10^{12.5}\, {\rm GeV}$ and $\sim 10^{11.75}-10^{12.25}\, {\rm GeV}$, respectively. 
Compared with the wider range adopted in figure~\ref{fig_ratio_stochastic_a}, where the spread in decay constants alone can make any axion dominant, the narrower ranges suppress this spread and thus isolate the effect of the hierarchical axion masses. 
In this way, the sequential mass crossings and the resulting preference for the heaviest ALPs become more transparent.
Note that in these figures, when considering axion mass contributions, we assume the mass approximations in eq.~\eqref{mass_condition} are valid for all ALPs.
In addition, although the ALP decay constants are considered within a narrow range --- which includes the QCD axion decay constant --- they must remain distinct from it and maintain a certain separation; otherwise, the classification of each ALP into the light or heavy regime and the hierarchical mass condition required for efficient energy transfer would become ambiguous.
This treatment aims to investigate the axion mass effects within the stochastic mixing mechanism, while preserving the general conclusion that once mass contributions are included, heavier ALPs tend to dominate after mixing.

\section{The role of ultralight axion-like particles}
\label{ultra_light_ALPs}    
 
In this section, we investigate the role of ultralight ALPs within the multi-component axion scenario, considering the stochastic mixing mechanism. 

We first consider the two-component framework, specifically a model comprising the QCD axion ($a_0$) and an ALP ($A$).
Note that since our goal is to explore the role of ultralight ALPs, and given that there is only one QCD axion, we normalize the numerical analysis to the QCD axion DM abundance. 
A comprehensive display of the total axion abundance is currently of secondary importance given the large number of model parameters. 
According to section~\ref{multi_component} and the stochastic mixing framework, in which each ALP may lie on either side of $f_{a0}$, the QCD axion DM abundance can be described by
\begin{eqnarray}
\begin{aligned}
\Omega_{a_0} h^2&\simeq0.14 \left(\dfrac{g_{*s}(T_0)}{3.94}\right)\left(\dfrac{g_*(T_{{\rm osc}, A})}{61.75}\right)^{-1/4}\\
&\times \left(\dfrac{f_{a_0}}{10^{7}\, \rm GeV}\right)^{-1} \left(\dfrac{f_A}{10^{11.2}\, \rm GeV}\right)^{2} \left(\dfrac{m_A}{10^{-2}\, \rm eV}\right)^{-1/2} \left\langle\Theta_i^2f(\Theta_i)\right\rangle\, ,
\label{QCD_a_abundance_1}
\end{aligned}
\end{eqnarray}
or
\begin{eqnarray}
\begin{aligned}
\Omega_{a_0} h^2&\simeq0.14 \left(\dfrac{g_{*s}(T_0)}{3.94}\right)\left(\dfrac{g_*(T_{{\rm osc}, A})}{61.75}\right)^{-1/4}\\
&\times \left(\dfrac{f_{a_0}}{10^{14}\, \rm GeV}\right)^{-1} \left(\dfrac{f_A}{10^{13.2}\, \rm GeV}\right)^{2} \left(\dfrac{m_A}{10^{-8}\, \rm eV}\right)^{-1/2} \left\langle\Theta_i^2f(\Theta_i)\right\rangle\, .
\label{QCD_a_abundance_2}
\end{aligned}
\end{eqnarray}
Here $f(\Theta_i)$ is an anharmonicity factor, and for simplicity, we set $\left\langle\Theta_i^2f(\Theta_i)\right\rangle\sim 1$.
Note that eqs.~\eqref{QCD_a_abundance_1} and \eqref{QCD_a_abundance_2} above correspond to two different cases in stochastic mixing: $f_A>f_{a_0}$ and $f_A<f_{a_0}$, respectively. 
Although written as two separate formulas, they have the same form and differ only
in the reference values chosen for the parameters.
For the sake of comparison, we set $f_{a_0}$ to be very small and very large, respectively. 
Furthermore, we set the ALP mass $m_A$ to be smaller than $m_{a_0,0}$.

\begin{figure}[t]
\centering
\includegraphics[width=0.7\textwidth]{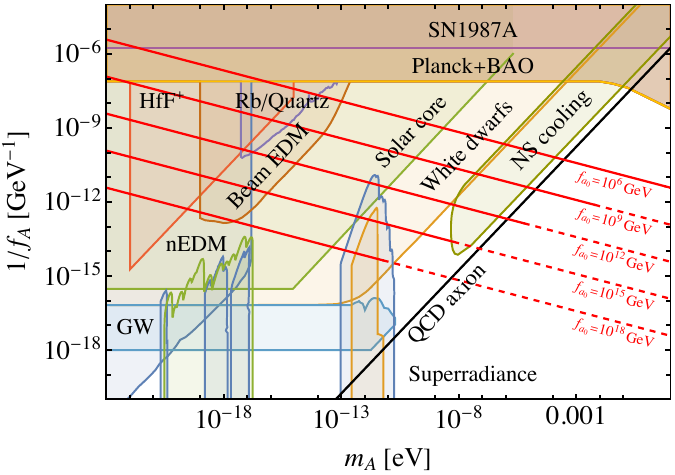}
\caption{Constraints on ALPs in the $\{m_A, 1/f_A\}$ plane for the two-component axion DM scenario with stochastic axion mixing. 
The thick red lines correspond to QCD axion decay constants ranging from $f_{a_0}=10^{6}\, \rm GeV$ to $10^{18}\, \rm GeV$. 
The solid and dashed red lines correspond to $m_A<m_{a_0,0}$ and $m_A>m_{a_0,0}$, respectively. 
Additional constraints on ALPs in this plane are adopted from ref.~\cite{ciaran_o_hare_2020_3932430}.}
\label{fig_mAfA}
\end{figure}  

In figure~\ref{fig_mAfA}, we present the ALP distribution in the $\{m_A, 1/f_A\}$ plane.
Each thick red line corresponds to a fixed QCD axion decay constant $f_{a_0}=10^{6}$--$10^{18}\, \rm GeV$, along which the QCD axion alone can account for the observed DM abundance for a suitable choice of the ALP parameters.
For a given $f_{a_0}$, the two branches of the red line correspond to $f_A>f_{a_0}$ and $f_A<f_{a_0}$, $\rm i.e.$, the two cases described by eqs.~\eqref{QCD_a_abundance_1} and \eqref{QCD_a_abundance_2}, respectively.
The solid portions of these red lines denote the regions where $m_A<m_{a_0,0}$, so that the ALP is lighter than the QCD axion and the hierarchical mass condition required for efficient mass crossing can be satisfied; the dashed portions, where $m_A>m_{a_0,0}$, are shown only for completeness and lie outside the mass ordering adopted in the stochastic mixing framework.
For completeness, we also incorporate the existing constraints on ALPs in this plane from ref.~\cite{ciaran_o_hare_2020_3932430}, providing a comprehensive overview of the allowed parameter space.
We find that in the stochastic axion mixing scenario, the parameter space for the QCD axion decay constant that allows the axion to constitute the entirety of DM is significantly enlarged.
This is primarily because, in stochastic mixing, the QCD axion abundance can be either suppressed or enhanced depending on the choice of different ALP parameters; by adjusting $f_A$ and $m_A$, one can compensate for a much wider range of $f_{a_0}$ than in the standard misalignment scenario.
Specifically, as $f_{a_0}$ increases, the corresponding upper limit on the ALP mass decreases, implying that the ALPs become lighter.
This trend is consistent with the fact that the zero-temperature QCD axion mass scales as $m_{a_0,0}\propto1/f_{a_0}$, so that a larger $f_{a_0}$ lowers the mass anchor below which the ALP spectrum is built.
Furthermore, there are certain constraints on the lower bound of the ALP mass.
However, since this bound is derived from the adiabatic conversion of axion energy and the majority of the parameter space is already excluded by other existing constraints depicted in figure~\ref{fig_mAfA}, we will not discuss it further. 
In contrast, regarding the upper bound of the ALP mass, most of the corresponding parameter space --- regardless of whether the QCD axion decay constant is extremely small or large --- remains unconstrained and is accessible to future ALP detection experiments.
This represents one of the primary novel cosmological consequences of the stochastic mixing mechanism.

\begin{figure}[t]
\centering
\includegraphics[width=0.7\textwidth]{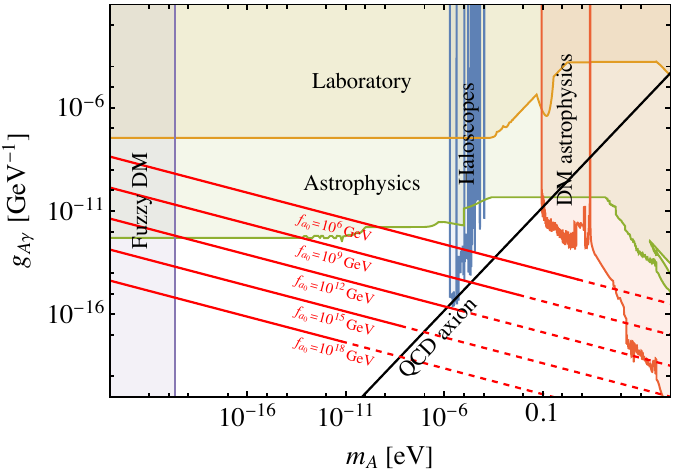}
\caption{Constraints on ALPs in the ALP-photon coupling $\{m_A, g_{A\gamma}\}$ plane for the two-component axion DM scenario with stochastic axion mixing. 
The thick red lines correspond to QCD axion decay constants ranging from $f_{a_0}=10^{6}\, \rm GeV$ to $10^{18}\, \rm GeV$. 
The solid and dashed red lines correspond to $m_A<m_{a_0,0}$ and $m_A>m_{a_0,0}$, respectively. 
Additional constraints on ALPs in this plane are adopted from ref.~\cite{ciaran_o_hare_2020_3932430}.}
\label{fig_mAgAg}
\end{figure} 

In addition, in figure~\ref{fig_mAgAg}, we also show the ALP distribution in the $\{m_A, g_{A\gamma}\}$ plane, where $g_{A\gamma}$ denotes the ALP-photon coupling constant. 
Note that
\begin{eqnarray}
g_{{a_0}\gamma} &=& \dfrac{\alpha}{2\pi} \dfrac{C_{{a_0}\gamma}}{f_{a_0}}\, ,\quad C_{{a_0}\gamma}= \dfrac{E}{N} - 1.92 \, ,\\
g_{A\gamma} &=& \dfrac{\alpha}{2\pi} \dfrac{C_{A\gamma}}{f_A}\, ,\quad C_{A\gamma}\simeq\mathcal{O}(1)\, ,
\end{eqnarray}
where $C_{{a_0}\gamma}$ and $C_{A\gamma}$ are model-dependent constants, $E/N=8/3$ for the DFSZ-I axion, $\alpha\simeq1/137$ is the fine-structure constant, and we take $C_{A\gamma}\simeq\mathcal{O}(1)$ for simplicity.
Since $g_{A\gamma}$ is proportional to the inverse of the ALP decay constant $f_A$, the resulting ALP distribution (red solid and dashed lines) is similar to that in figure~\ref{fig_mAfA}; the solid and dashed portions again correspond to $m_A<m_{a_0,0}$ and $m_A>m_{a_0,0}$, respectively.
In this plane, each red line is obtained by converting the corresponding $f_A$ in figure~\ref{fig_mAfA} through $g_{A\gamma}\propto1/f_A$, so that a larger $f_A$ corresponds to a smaller $g_{A\gamma}$ and vice versa.
We note that $C_{A\gamma}\simeq\mathcal{O}(1)$ is a simplifying assumption; in specific ultraviolet completions, $C_{A\gamma}$ can differ by an $\mathcal{O}(1)$ factor, which would shift the red lines vertically without changing their shape or the qualitative conclusions.
The ALP-photon coupling is one of the most extensively searched for ALP interactions, and the corresponding constraints are among the strongest for a wide range of ALP masses.
However, under current ALP-photon coupling constraints, the bounds on light ALPs remain relatively weak, and most light ALPs have not yet been excluded.
This is particularly relevant for ultralight ALPs with $m_A\lesssim10^{-11}\,\rm eV$, whose oscillation frequencies lie below the reach of conventional cavity haloscopes but can be probed by lumped-element detectors, low-frequency resonators, and astronomical observations.
Consequently, the light ALP region of interest is likely to be probed or excluded by upcoming ALP-photon coupling experiments in the near future, particularly through astronomical observations and indirect DM searches. 
Although some lower bounds on the ALP mass may have been set, this parameter space remains sufficiently large. 

\begin{figure}[t]
\centering
\includegraphics[width=0.7\textwidth]{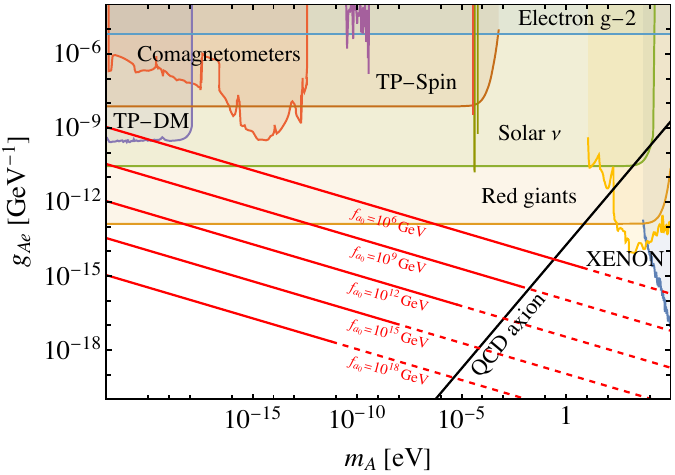}
\caption{Same as figure~\ref{fig_mAgAg} but for the ALP-electron coupling $\{m_A, g_{Ae}\}$ plane.
The solid and dashed red lines correspond to $m_A<m_{a_0,0}$ and $m_A>m_{a_0,0}$, respectively. 
Additional constraints on ALPs in this plane are adopted from ref.~\cite{ciaran_o_hare_2020_3932430}.}
\label{fig_mAgAe}
\end{figure}  

\begin{figure}[t]
\centering
\includegraphics[width=0.7\textwidth]{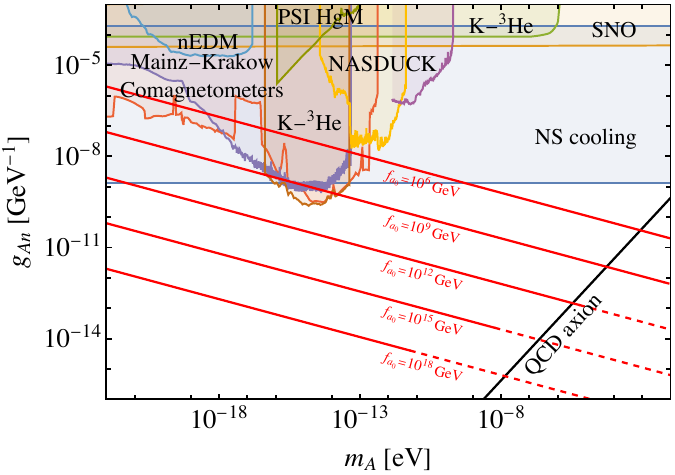}
\caption{Same as figure~\ref{fig_mAgAg} but for the ALP-neutron coupling $\{m_A, g_{An}\}$ plane.
The solid and dashed red lines correspond to $m_A<m_{a_0,0}$ and $m_A>m_{a_0,0}$, respectively. 
Additional constraints on ALPs in this plane are adopted from ref.~\cite{ciaran_o_hare_2020_3932430}.}
\label{fig_mAgAn}
\end{figure}

\begin{figure}[t]
\centering
\includegraphics[width=0.7\textwidth]{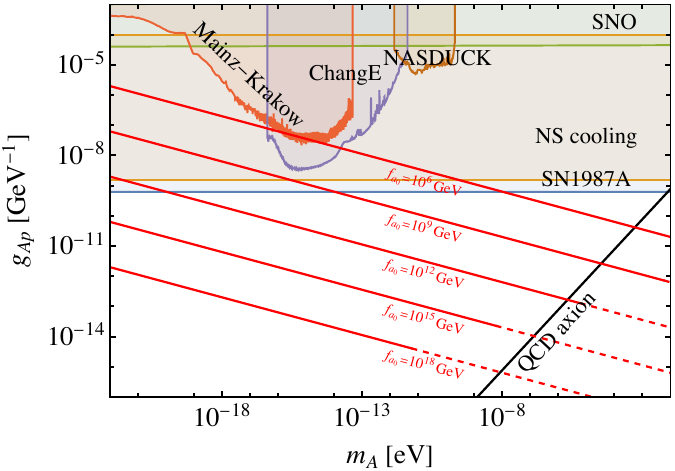}
\caption{Same as figure~\ref{fig_mAgAg} but for the ALP-proton coupling $\{m_A, g_{Ap}\}$ plane.
The solid and dashed red lines correspond to $m_A<m_{a_0,0}$ and $m_A>m_{a_0,0}$, respectively. 
Additional constraints on ALPs in this plane are adopted from ref.~\cite{ciaran_o_hare_2020_3932430}.}
\label{fig_mAgAp}
\end{figure}

Similar conclusions hold for the ALP couplings to electrons ($e$), neutrons ($n$), and protons ($p$), as shown in figures~\ref{fig_mAgAe}, \ref{fig_mAgAn}, and \ref{fig_mAgAp}, respectively. 
The red lines in these figures are obtained in the same way by converting $f_A$ through $g_{Ae}\propto1/f_A$, $g_{An}\propto1/f_A$, and $g_{Ap}\propto1/f_A$, respectively.
The corresponding couplings are given by
\begin{eqnarray}
g_{{a_0}e} &=& \dfrac{C_{{a_0}e}m_e}{f_{a_0}}\, ,\quad C_{{a_0}e}= 2\times 10^{-4} \, ({\rm KSVZ})\, ,\\
g_{Ae} &=& \dfrac{C_{Ae}m_e}{f_A}\, ,\quad C_{Ae}\simeq\mathcal{O}(1)\, ,\\
g_{{a_0}n} &=& \dfrac{C_{{a_0}n}m_n}{f_{a_0}}\, ,\quad C_{{a_0}n}= 0.26 \, ({\rm DFSZ})\, ,\\
g_{An} &=& \dfrac{C_{An}m_n}{f_A}\, ,\quad C_{An}\simeq\mathcal{O}(1)\, ,\\
g_{{a_0}p} &=& \dfrac{C_{{a_0}p}m_p}{f_{a_0}}\, ,\quad C_{{a_0}p}= -0.47 \, ({\rm KSVZ})\, ,\\
g_{Ap} &=& \dfrac{C_{Ap}m_p}{f_A}\, ,\quad C_{Ap}\simeq\mathcal{O}(1)\, ,
\end{eqnarray}
where $m_e\simeq 5.11\times10^{-4}\, \rm GeV$, $m_n\simeq 0.939565\, \rm GeV$, and $m_p\simeq 0.938272\, \rm GeV$ are the masses of the electron, neutron, and proton, respectively.
Similarly, for simplicity, we take $C_{Ae}, C_{An}, C_{Ap}\simeq\mathcal{O}(1)$.
Note that in figures~\ref{fig_mAgAn} (for neutrons) and \ref{fig_mAgAp} (for protons), some regions with $m_A>m_{a_0,0}$ (dashed lines) are not shown because the existing constraints for this part are not yet clear.

\begin{figure}[t]
\centering
\includegraphics[width=0.7\textwidth]{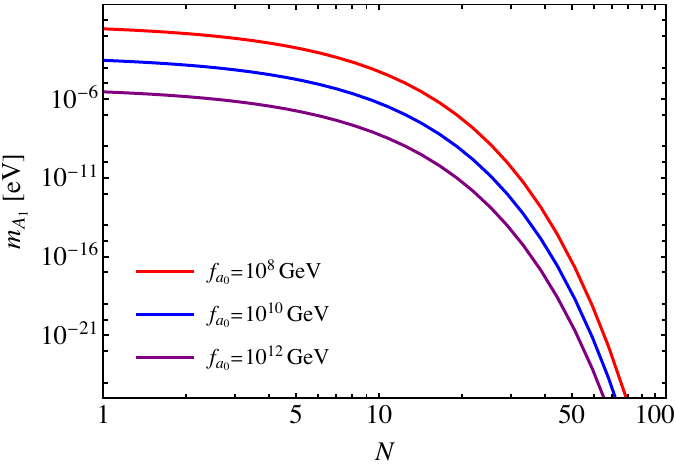}
\caption{Upper limits on the mass of the lightest ALP $A_1$ in stochastic axion mixing as a function of the total number of ALPs $N$.
The lines follow from the hierarchical mass condition in eq.~\eqref{mass_condition}, $\rm i.e.$, $m_{A_1}\lesssim 2^{-N}m_{a_0,0}$, assuming the ALP masses increase monotonically with $i$. 
The red, blue, and purple lines represent $f_{a_0}=10^{8}\, \rm GeV$, $10^{10}\, \rm GeV$, and $10^{12}\, \rm GeV$, respectively.}
\label{fig_N-mA}
\end{figure}

Next, we consider the multi-component scenario comprising the QCD axion ($a_0$) and $N$ ALPs ($A_i$), with the mass ordering $m_{A_i}<m_{A_{i+1}}<m_{a_0,0}$. 
Figure~\ref{fig_N-mA} presents the upper limits on the mass of the lightest ALP $A_1$ in stochastic axion mixing as a function of the total number of ALPs, $N$. 
The lines are not obtained from a scan over the individual ALP decay constants $f_{A_i}$; rather, they follow from the hierarchical mass condition in eq.~\eqref{mass_condition}, which is required for efficient mass crossing and energy transfer in the stochastic mixing framework. 
For the lightest ALP, this condition implies approximately
\begin{equation}
m_{A_1}\lesssim 2^{-N}m_{a_0,0}\, ,
\end{equation}
up to $O(1)$ factors and anharmonic corrections. 
Thus, for each fixed $f_{a_0}$ and $N$, the corresponding line in figure~\ref{fig_N-mA} represents the maximal $m_{A_1}$ compatible with the hierarchical condition. 
Equivalently, it is the envelope allowed by the requirement that $N$ ALPs can be accommodated below the zero-temperature QCD axion mass with successive mass ratios of order $1/2$ or smaller.

Several features of figure~\ref{fig_N-mA} are worth emphasizing. 
First, the upper mass limit of $A_1$ decreases rapidly with increasing $N$. 
This behavior is essentially exponential in $N$, as expected from the $2^{-N}$ scaling in the hierarchical condition, and becomes particularly pronounced for $N\sim\mathcal{O}(100)$, where the lightest ALP is pushed into the extremely ultralight regime. 
Second, for a fixed $N$, the upper limit on $m_{A_1}$ decreases as the QCD axion decay constant $f_{a_0}$ increases. 
This is primarily because the zero-temperature QCD axion mass scales as $m_{a_0,0}\propto 1/f_{a_0}$; a larger $f_{a_0}$ therefore lowers the overall mass anchor from which the hierarchical ALP spectrum is built. 
The red, blue, and purple lines in figure~\ref{fig_N-mA}, corresponding to $f_{a_0}=10^{8}\,\mathrm{GeV}$, $10^{10}\,\mathrm{GeV}$, and $10^{12}\,\mathrm{GeV}$, respectively, clearly exhibit this trend.

The above scaling is physically relevant for string axiverse constructions. 
In the type IIB string axiverse, the presence of $\sim\mathcal{O}(100)$ light ALPs with hierarchical masses and decay constants is generic and naturally includes the QCD axion \cite{Cicoli:2012sz}. 
In such a spectrum, a large number of ALPs do not simply populate the same mass window; rather, they force the lightest state to be exponentially lighter if the hierarchical condition is to be satisfied. 
Moreover, in stochastic mixing, the decay constants of these ultralight ALPs ($f_{A_i}$) are generally not restricted to be small and can take on very large values. 
This leads to a key conclusion:
\begin{itemize}
\item The larger the population of axions/ALPs in the stochastic mixing model, the more ultralight ALPs emerge, thus allowing for a greater number of viable ALPs with large decay constants.
\end{itemize}
These ALPs, which currently evade experimental bounds, represent a key focus for future axion detection experiments. 
We caution that the lines in figure~\ref{fig_N-mA} should be regarded as the representative upper envelope obtained from eq.~\eqref{mass_condition}; for decay constants spanning a wider range or for non-universal $O(1)$ coefficients, the detailed upper bound may shift by $O(1)$ factors, but the exponential dependence on $N$ and the monotonic dependence on $f_{a_0}$ remain robust.

\section{Conclusion}
\label{sec_Conclusion}

In summary, we have revisited the multi-component axion DM scenario and explored the role of ultralight ALPs within this framework by incorporating the recently proposed stochastic axion mixing mechanism. 
Compared with the maximal mixing case, the stochastic domain wall matrix provides a unified parametrization of the light and heavy axion regimes, with each ALP classified according to whether its decay constant lies below or above that of the QCD axion. 
This generality allows the post-mixing energy density to be analyzed in a more model-independent way, while still requiring that the ALP masses remain below the zero-temperature QCD axion mass and be non-degenerate for efficient mass crossing.
Consequently, the resulting DM phenomenology is richer than a simple interpolation between the light and heavy axion limits.

Our analysis reveals several key insights. First, unlike scenarios with maximal mixing, the stochastic mechanism allows, depending on the parameters, either the QCD axion or any of the ALPs to dominate the post-mixing energy density. 
In the absence of axion mass contributions, the post-mixing abundance is controlled only by the decay constants, so the dominant state is not necessarily the QCD axion and can be any one of the ALPs. 
Once axion mass contributions are included, the sequential mass crossings redistribute the energy density according to the mass ordering, and the ratio $\rho_{A_i,0}/\rho_{a_0,0}$ increases with the ALP mass, so that the heaviest ALPs tend to prevail. 
This behavior is most transparent when the ALP decay constants are drawn from a narrow range around $f_{a_0}$, where the hierarchical mass condition required for efficient energy transfer can be satisfied. 
For a wider or strongly non-universal distribution of $f_{A_i}$, the identity of the dominant component may depend on the detailed competition between decay constants and masses, although the general trend that heavier ALPs are favored after mass contributions are included remains valid.

Second, within a two-component framework, stochastic mixing significantly expands the viable parameter space for the QCD axion decay constant, $f_{a_0}$, allowing the QCD axion to constitute the entirety of the observed cold DM abundance even for values of $f_{a_0}$ that would typically lead to overproduction or underproduction. 
The corresponding ALP mass range remains largely unconstrained by current experiments, and for larger $f_{a_0}$ the upper limit on the ALP mass shifts to lighter values.
These regions are therefore promising targets for future axion searches, especially through ALP-photon, ALP-electron, and ALP-nucleon couplings.

Finally, extending our analysis to the multi-component case, we find that a larger population of ALPs leads to the emergence of more ultralight ALPs with large decay constants. 
In the stochastic mixing framework, the hierarchical mass condition implies $m_{A_1}\lesssim 2^{-N}m_{a_0,0}$, so that the lightest ALP is pushed into the extremely ultralight regime as $N$ increases; for fixed $N$, a larger $f_{a_0}$ further lowers this upper bound. 
Such a spectrum is physically well motivated in string axiverse constructions, where $\mathcal{O}(100)$ light ALPs with hierarchical masses and decay constants can naturally arise together
with the QCD axion. 
We caution that the corresponding lines should be regarded as representative upper envelopes; for non-universal $O(1)$ coefficients or wider decay constant ranges, the detailed bounds may shift, but the exponential dependence on $N$ and the monotonic dependence on $f_{a_0}$ remain robust. 
These ultralight ALPs, which currently evade experimental constraints, represent compelling targets for upcoming axion searches. 
Our work underscores the rich phenomenology of multi-component axion DM and highlights the crucial role of ultralight ALPs as a bridge between theoretical models and future observational probes.

\section*{Acknowledgments}

This work was supported by Yunnan University and the Key Laboratory of Astroparticle Physics of Yunnan Province.

\appendix



\bibliographystyle{JHEP}
\bibliography{references}

\providecommand{\href}[2]{#2}\begingroup\raggedright\begin{thebibliography}{10}

\bibitem{Cirelli:2024ssz}
M.~Cirelli, A.~Strumia and J.~Zupan, \emph{{Dark Matter}},
  \href{http://arxiv.org/abs/2406.01705}{{\tt 2406.01705}}.

\bibitem{Peccei:1977hh}
R.~Peccei and H.~R. Quinn, \emph{{CP Conservation in the Presence of
  Instantons}},
  \href{http://dx.doi.org/10.1103/PhysRevLett.38.1440}{\emph{Phys. Rev. Lett.}
  {\bf 38} (1977) 1440--1443}.

\bibitem{Peccei:1977ur}
R.~Peccei and H.~R. Quinn, \emph{{Constraints Imposed by CP Conservation in the
  Presence of Instantons}},
  \href{http://dx.doi.org/10.1103/PhysRevD.16.1791}{\emph{Phys. Rev. D} {\bf
  16} (1977) 1791--1797}.

\bibitem{Weinberg:1977ma}
S.~Weinberg, \emph{{A New Light Boson?}},
  \href{http://dx.doi.org/10.1103/PhysRevLett.40.223}{\emph{Phys. Rev. Lett.}
  {\bf 40} (1978) 223--226}.

\bibitem{Wilczek:1977pj}
F.~Wilczek, \emph{{Problem of Strong $P$ and $T$ Invariance in the Presence of
  Instantons}}, \href{http://dx.doi.org/10.1103/PhysRevLett.40.279}{\emph{Phys.
  Rev. Lett.} {\bf 40} (1978) 279--282}.

\bibitem{Kim:1979if}
J.~E. Kim, \emph{{Weak Interaction Singlet and Strong CP Invariance}},
  \href{http://dx.doi.org/10.1103/PhysRevLett.43.103}{\emph{Phys. Rev. Lett.}
  {\bf 43} (1979) 103}.

\bibitem{Shifman:1979if}
M.~A. Shifman, A.~I. Vainshtein and V.~I. Zakharov, \emph{{Can Confinement
  Ensure Natural CP Invariance of Strong Interactions?}},
  \href{http://dx.doi.org/10.1016/0550-3213(80)90209-6}{\emph{Nucl. Phys. B}
  {\bf 166} (1980) 493--506}.

\bibitem{Dine:1981rt}
M.~Dine, W.~Fischler and M.~Srednicki, \emph{{A Simple Solution to the Strong
  CP Problem with a Harmless Axion}},
  \href{http://dx.doi.org/10.1016/0370-2693(81)90590-6}{\emph{Phys. Lett. B}
  {\bf 104} (1981) 199--202}.

\bibitem{Zhitnitsky:1980tq}
A.~R. Zhitnitsky, \emph{{On Possible Suppression of the Axion Hadron
  Interactions. (In Russian)}}, {\emph{Sov. J. Nucl. Phys.} {\bf 31} (1980)
  260}.

\bibitem{Witten:1984dg}
E.~Witten, \emph{{Some Properties of O(32) Superstrings}},
  \href{http://dx.doi.org/10.1016/0370-2693(84)90422-2}{\emph{Phys. Lett. B}
  {\bf 149} (1984) 351--356}.

\bibitem{Green:1984sg}
M.~B. Green and J.~H. Schwarz, \emph{{Anomaly Cancellation in Supersymmetric
  D=10 Gauge Theory and Superstring Theory}},
  \href{http://dx.doi.org/10.1016/0370-2693(84)91565-X}{\emph{Phys. Lett. B}
  {\bf 149} (1984) 117--122}.

\bibitem{Choi:2003wr}
K.-w. Choi, \emph{{A QCD axion from higher dimensional gauge field}},
  \href{http://dx.doi.org/10.1103/PhysRevLett.92.101602}{\emph{Phys. Rev.
  Lett.} {\bf 92} (2004) 101602},
  [\href{http://arxiv.org/abs/hep-ph/0308024}{{\tt hep-ph/0308024}}].

\bibitem{Reece:2024wrn}
M.~Reece, \emph{{Extra-dimensional axion expectations}},
  \href{http://dx.doi.org/10.1007/JHEP07(2025)130}{\emph{JHEP} {\bf 07} (2025)
  130}, [\href{http://arxiv.org/abs/2406.08543}{{\tt 2406.08543}}].

\bibitem{Svrcek:2006yi}
P.~Svrcek and E.~Witten, \emph{{Axions In String Theory}},
  \href{http://dx.doi.org/10.1088/1126-6708/2006/06/051}{\emph{JHEP} {\bf 06}
  (2006) 051}, [\href{http://arxiv.org/abs/hep-th/0605206}{{\tt
  hep-th/0605206}}].

\bibitem{Conlon:2006tq}
J.~P. Conlon, \emph{{The QCD axion and moduli stabilisation}},
  \href{http://dx.doi.org/10.1088/1126-6708/2006/05/078}{\emph{JHEP} {\bf 05}
  (2006) 078}, [\href{http://arxiv.org/abs/hep-th/0602233}{{\tt
  hep-th/0602233}}].

\bibitem{Arvanitaki:2009fg}
A.~Arvanitaki, S.~Dimopoulos, S.~Dubovsky, N.~Kaloper and J.~March-Russell,
  \emph{{String Axiverse}},
  \href{http://dx.doi.org/10.1103/PhysRevD.81.123530}{\emph{Phys. Rev. D} {\bf
  81} (2010) 123530}, [\href{http://arxiv.org/abs/0905.4720}{{\tt 0905.4720}}].

\bibitem{Acharya:2010zx}
B.~S. Acharya, K.~Bobkov and P.~Kumar, \emph{{An M Theory Solution to the
  Strong CP Problem and Constraints on the Axiverse}},
  \href{http://dx.doi.org/10.1007/JHEP11(2010)105}{\emph{JHEP} {\bf 11} (2010)
  105}, [\href{http://arxiv.org/abs/1004.5138}{{\tt 1004.5138}}].

\bibitem{Cicoli:2012sz}
M.~Cicoli, M.~Goodsell and A.~Ringwald, \emph{{The type IIB string axiverse and
  its low-energy phenomenology}},
  \href{http://dx.doi.org/10.1007/JHEP10(2012)146}{\emph{JHEP} {\bf 10} (2012)
  146}, [\href{http://arxiv.org/abs/1206.0819}{{\tt 1206.0819}}].

\bibitem{Preskill:1982cy}
J.~Preskill, M.~B. Wise and F.~Wilczek, \emph{{Cosmology of the Invisible
  Axion}}, \href{http://dx.doi.org/10.1016/0370-2693(83)90637-8}{\emph{Phys.
  Lett. B} {\bf 120} (1983) 127--132}.

\bibitem{Abbott:1982af}
L.~Abbott and P.~Sikivie, \emph{{A Cosmological Bound on the Invisible Axion}},
  \href{http://dx.doi.org/10.1016/0370-2693(83)90638-X}{\emph{Phys. Lett. B}
  {\bf 120} (1983) 133--136}.

\bibitem{Dine:1982ah}
M.~Dine and W.~Fischler, \emph{{The Not So Harmless Axion}},
  \href{http://dx.doi.org/10.1016/0370-2693(83)90639-1}{\emph{Phys. Lett. B}
  {\bf 120} (1983) 137--141}.

\bibitem{Marsh:2015xka}
D.~J.~E. Marsh, \emph{{Axion Cosmology}},
  \href{http://dx.doi.org/10.1016/j.physrep.2016.06.005}{\emph{Phys. Rept.}
  {\bf 643} (2016) 1--79}, [\href{http://arxiv.org/abs/1510.07633}{{\tt
  1510.07633}}].

\bibitem{OHare:2024nmr}
C.~A.~J. O'Hare, \emph{{Cosmology of axion dark matter}},
  \href{http://dx.doi.org/10.22323/1.454.0040}{\emph{PoS} {\bf COSMICWISPers}
  (2024) 040}, [\href{http://arxiv.org/abs/2403.17697}{{\tt 2403.17697}}].

\bibitem{Montero:2022prj}
M.~Montero, C.~Vafa and I.~Valenzuela, \emph{{The dark dimension and the
  Swampland}}, \href{http://dx.doi.org/10.1007/JHEP02(2023)022}{\emph{JHEP}
  {\bf 02} (2023) 022}, [\href{http://arxiv.org/abs/2205.12293}{{\tt
  2205.12293}}].

\bibitem{Gendler:2024gdo}
N.~Gendler and C.~Vafa, \emph{{Axions in the dark dimension}},
  \href{http://dx.doi.org/10.1007/JHEP12(2024)127}{\emph{JHEP} {\bf 12} (2024)
  127}, [\href{http://arxiv.org/abs/2404.15414}{{\tt 2404.15414}}].

\bibitem{Li:2024jko}
H.-J. Li, \emph{{QCD axion dark matter in the dark dimension}},
  \href{http://dx.doi.org/10.1007/JHEP05(2025)139}{\emph{JHEP} {\bf 05} (2025)
  139}, [\href{http://arxiv.org/abs/2412.19426}{{\tt 2412.19426}}].

\bibitem{Daido:2015cba}
R.~Daido, N.~Kitajima and F.~Takahashi, \emph{{Level crossing between the QCD
  axion and an axionlike particle}},
  \href{http://dx.doi.org/10.1103/PhysRevD.93.075027}{\emph{Phys. Rev. D} {\bf
  93} (2016) 075027}, [\href{http://arxiv.org/abs/1510.06675}{{\tt
  1510.06675}}].

\bibitem{Li:2023uvt}
H.-J. Li, Y.-Q. Peng, W.~Chao and Y.-F. Zhou, \emph{{Light QCD axion dark
  matter from double level crossings}},
  \href{http://dx.doi.org/10.1016/j.physletb.2024.138444}{\emph{Phys. Lett. B}
  {\bf 849} (2024) 138444}, [\href{http://arxiv.org/abs/2310.02126}{{\tt
  2310.02126}}].

\bibitem{Cyncynates:2023esj}
D.~Cyncynates and J.~O. Thompson, \emph{{Heavy QCD axion dark matter from
  avoided level crossing}},
  \href{http://dx.doi.org/10.1103/PhysRevD.108.L091703}{\emph{Phys. Rev. D}
  {\bf 108} (2023) L091703}, [\href{http://arxiv.org/abs/2306.04678}{{\tt
  2306.04678}}].

\bibitem{Hill:1988bu}
C.~T. Hill and G.~G. Ross, \emph{{Models and New Phenomenological Implications
  of a Class of Pseudogoldstone Bosons}},
  \href{http://dx.doi.org/10.1016/0550-3213(88)90062-4}{\emph{Nucl. Phys. B}
  {\bf 311} (1988) 253--297}.

\bibitem{Ho:2018qur}
S.-Y. Ho, K.~Saikawa and F.~Takahashi, \emph{{Enhanced photon coupling of ALP
  dark matter adiabatically converted from the QCD axion}},
  \href{http://dx.doi.org/10.1088/1475-7516/2018/10/042}{\emph{JCAP} {\bf 10}
  (2018) 042}, [\href{http://arxiv.org/abs/1806.09551}{{\tt 1806.09551}}].

\bibitem{Li:2023xkn}
H.-J. Li, \emph{{Axion dark matter with explicit Peccei-Quinn symmetry breaking
  in the axiverse}},
  \href{http://dx.doi.org/10.1088/1475-7516/2024/09/025}{\emph{JCAP} {\bf 09}
  (2024) 025}, [\href{http://arxiv.org/abs/2307.09245}{{\tt 2307.09245}}].

\bibitem{Li:2025imm}
H.-J. Li, W.~Chao, H.-K. Guo and Y.-F. Zhou, \emph{{Completing axion double
  level crossings}},
  \href{http://dx.doi.org/10.1016/j.nuclphysb.2026.117455}{\emph{Nucl. Phys. B}
  {\bf 1026} (2026) 117455}, [\href{http://arxiv.org/abs/2508.12722}{{\tt
  2508.12722}}].

\bibitem{Murai:2024nsp}
K.~Murai, Y.~Narita, F.~Takahashi and W.~Yin, \emph{{QCD axion dark matter from
  level crossing with refined adiabatic condition}},
  \href{http://dx.doi.org/10.1007/JHEP04(2025)124}{\emph{JHEP} {\bf 04} (2025)
  124}, [\href{http://arxiv.org/abs/2412.10232}{{\tt 2412.10232}}].

\bibitem{Li:2025uwq}
H.-J. Li, \emph{{Some properties of multi-component axion dark matter}},
  \href{http://dx.doi.org/10.1016/j.physletb.2025.140141}{\emph{Phys. Lett. B}
  {\bf 873} (2026) 140141}, [\href{http://arxiv.org/abs/2510.22569}{{\tt
  2510.22569}}].

\bibitem{Murai:2025wbg}
K.~Murai, F.~Takahashi and K.~Tomiyama, \emph{{Sign-flipping axion potentials
  via Kapitza-type modulation by heavy axions}},
  \href{http://dx.doi.org/10.1007/JHEP07(2026)171}{\emph{JHEP} {\bf 07} (2026)
  171}, [\href{http://arxiv.org/abs/2509.11032}{{\tt 2509.11032}}].

\bibitem{Li:2025cep}
H.-J. Li and Y.-F. Zhou, \emph{{Axion mixing in the string axiverse}},
  \href{http://dx.doi.org/10.1016/j.nuclphysb.2026.117574}{\emph{Nucl. Phys. B}
  {\bf 1029} (2026) 117574}, [\href{http://arxiv.org/abs/2504.10170}{{\tt
  2504.10170}}].

\bibitem{Li:2026rga}
H.-J. Li, \emph{{Stochastic Axion Mixing: A General Mechanism Beyond Decay
  Constant Constraints}},  \href{http://arxiv.org/abs/2604.27784}{{\tt
  2604.27784}}.

\bibitem{Asadi:2025cvm}
P.~Asadi, D.~Cyncynates and S.~Gori, \emph{{Axiverse Baryogenesis}},
  \href{http://arxiv.org/abs/2511.15794}{{\tt 2511.15794}}.

\bibitem{deGiorgi:2025ldc}
A.~de~Giorgi, J.~Jaeckel, S.~Monath and V.~Takhistov, \emph{{Multiple Axions in
  Laboratory Experiments}},  \href{http://arxiv.org/abs/2512.16837}{{\tt
  2512.16837}}.

\bibitem{Baryakhtar:2026oun}
M.~Baryakhtar, D.~Cyncynates and E.~Henry, \emph{{Axiverse Lampposts}},
  \href{http://arxiv.org/abs/2602.23424}{{\tt 2602.23424}}.

\bibitem{FernandezNavarro:2026cyu}
M.~Fern{\'a}ndez~Navarro, M.~F. Zamoro, M.~Pesut and X.~Ponce~D{\'\i}az,
  \emph{{The structure of multi-axion solutions to the strong CP problem}},
  \href{http://arxiv.org/abs/2605.06787}{{\tt 2605.06787}}.

\bibitem{Gorgulho:2026psp}
D.~S. Gorgulho, J.~A. Litterer and J.~G. Rosa, \emph{{String Axiverse
  Enhancement of Superradiant Dark Matter Production}},
  \href{http://arxiv.org/abs/2606.20524}{{\tt 2606.20524}}.

\bibitem{Langhoff:2026qqg}
K.~Langhoff, M.~Ramos and M.~Reig, \emph{{The Dark Dimension meets the
  Axiverse}},  \href{http://arxiv.org/abs/2607.27314}{{\tt 2607.27314}}.

\bibitem{Sengupta:2026gyg}
A.~Sengupta, D.~Stojkovic and A.~G. Riess, \emph{{A Multi-Axion Ladder Across
  Cosmic History: From Inflation, BBN, and Early Dark Energy to Late-Time
  Accelerated Expansion}},  \href{http://arxiv.org/abs/2609.11907}{{\tt
  2609.11907}}.

\bibitem{Lei:2026cbn}
Z.-Y. Lei and J.-W. Cui, \emph{{Axion identification in multi-phase models}},
  \href{http://arxiv.org/abs/2610.24028}{{\tt 2610.24028}}.

\bibitem{Li:2025tdi}
H.-J. Li, \emph{{Toward the Effective Light and Heavy QCD Axion Scenarios}},
  \href{http://arxiv.org/abs/2512.22829}{{\tt 2512.22829}}.

\bibitem{ciaran_o_hare_2020_3932430}
C.~O'HARE, \emph{cajohare/axionlimits: Axionlimits},  July, 2020.
\newblock 10.5281/zenodo.3932430.

\end{thebibliography}\endgroup

\end{document}